\documentclass[fleqn,10pt]{wlscirep}
\usepackage[utf8]{inputenc}
\usepackage[T1]{fontenc}
\usepackage{amsmath}
\usepackage{xcolor}

\title{Design of a system for controlling a levitating sphere in superfluid $^3$He at extremely low temperatures}

\author[1,*]{Manuel Array\'as}
\author[1]{Jos\'e L. Trueba}
\author[1]{Carlos Uriarte}
\author[2]{Dmitry E. Zmeev}
\affil[1]{\'Area de Electromagnetismo, Universidad Rey Juan Carlos, Tulip\'
an s/n, 28933 M\'ostoles, Madrid, Spain.}
\affil[2]{Department of Physics, Lancaster University, Lancaster, LA1 4YB, UK.}

\affil[*]{manuel.arrayas@urjc.es}

\begin{abstract}
  We present a new mechanical probe to study the properties of superfluid $^3$He at microkelvin temperatures down to 100$\mu$K. The setup consists of a set of coils for levitating a superconducting sphere and controlling its motion in a wide variety of regimes. In particular, the realisation of motion of a levitating body at a uniform velocity presents both an experimental challenge and a promising direction into the study of the edge states in topological superfluid $^3$He-B.  We include the theoretical study of the device stability and simulations to illustrate the capabilities of the control system.   
\end{abstract}
\begin{document}

\flushbottom
\maketitle
%
%
\thispagestyle{empty}

\section*{Introduction}

Starting from Andronikashvili's experiment with the torsional pendulum \cite{Andronikashvili46}, the measurement of drag forces on oscillating mechanical resonators immersed in liquid helium has been one of the main tools in the study of its superfluid properties. On resonance, the drag force is equal and opposite in sign to the driving force, applied to the oscillator, and is known to the experimenter. Multiple properties of the superfluid can be successfully inferred from the behaviour of the drag force.
On the other hand, oscillatory motion necessarily imposes significant limits on the experimental capabilities. Certain phenomena only manifest when a mechanical probe moves uniformly in the superfluid and cannot be observed using oscillators. Examples include generation of homogeneous and isotropic turbulence produced by a tightly-fitted grid towed through a channel \cite{Liu07, Zmeev14, Zmeev15} and observation of superfluid currents beyond the Landau velocity \cite{Bradley16}. 

Despite being the simplest form of motion, uniform motion is notoriously difficult to implement in a low-temperature experiment, mainly due to the limitations on heat release and available space \cite{Liu07, Zmeev14, Bradley11}.
Another obstacle to realisation of a successful experiment with a uniformly moving body is that the drag force acting upon it is a lot more difficult to measure than the drag force on an oscillator in resonance \cite{Bradley11}.

Recently, the uniform motion technique was used to discover anomalously long  lifetimes of quasiparticles bound to the surfaces of superfluid $^3$He-B\cite{Autti20}. This time was found to be about 6\,ms, much longer than any other time scales in the system\cite{Kuorelahti18}. The discovery of the long lifetime of the quasiparticles also explains another recently observed anomalous phenomenon \--- exceptionally high apparent thermal conductivity in narrow channels of superfluid $^3$He: most likely it is the long-lived surface quasiparticles that carry significant energy along the walls of the channel\cite{Lotnyk20}. In the experiments\cite{Bradley16, Autti20}, the surface-bound quasiparticles could be promoted into bulk by accelerating a goalpost-shaped superconducting wire, with the crossbar moving parallel to itself. In this case different parts of the posts moved at different velocities. Also, the crossbar was necessarily connected to a large `bath' of surface-bound excitations on the walls of the experimental container and the states emptied into bulk by acceleration could be refilled from this `bath' by diffusion. A more favourable topology and well-defined geometry are required for more quantitative experiments. One such experiment is determination of the spectrum of surface excitations, potentially containing Majorana-like fermions\cite{Mizushima16}, by accelerating the body to various velocities. An example of a probe suitable for these experiments is a levitated sphere.

In this work we present the design of an experimental setup consisting of a levitation and control system for moving a body in a wide range of motion regimes, from oscillatory to uniform, inside superfluid $^3$He at temperatures down to 100\,$\mu$K. We have made the design to work for a superconducting spherical object made of indium. It is important to note that a similar experiment has been already performed in $^4$He, with a magnetic sphere being levitated between the plates of a superconducting niobium capacitor \cite{Shoepe95}. A large voltage was applied in order to charge the sphere, before the niobium was cooled down below its critical temperature. Then the DC voltage was switched off and oscillatory motion was induced by applying an oscillatory voltage amplitude up to a few volts. This arrangement, however, is not suitable for superfluid $^3$He experiments, mainly due to the high magnetic field in the experimental region during precooling before nuclear demagnetisation: this field then will be trapped by the superconductor. Moreover one of our main goals is to be able to move the body with a uniform velocity.     

Our levitating system comprises a set of superconducting coils. The main two concentric rings at the bottom of the cell filled with $^3$He produce the levitating force, which cancels gravity. The control system is provided by a set of lateral coils, which allow the control of the motion in the horizontal plane. In order to enable the control, the system must sit in the stability region of the magnetic and gravitational energies, which imposes some strict conditions on the final design. Our results prove that we can meet these conditions in a range of currents and masses of the objects which can levitate inside the experimental cell. In this work we provide such analysis of the stability conditions and perform simulations of the motion of the sphere in the superfluid.

\section*{Results}

\subsection*{The setup}
\begin{figure}
  \begin{center}
  \includegraphics[width=0.5\textwidth]{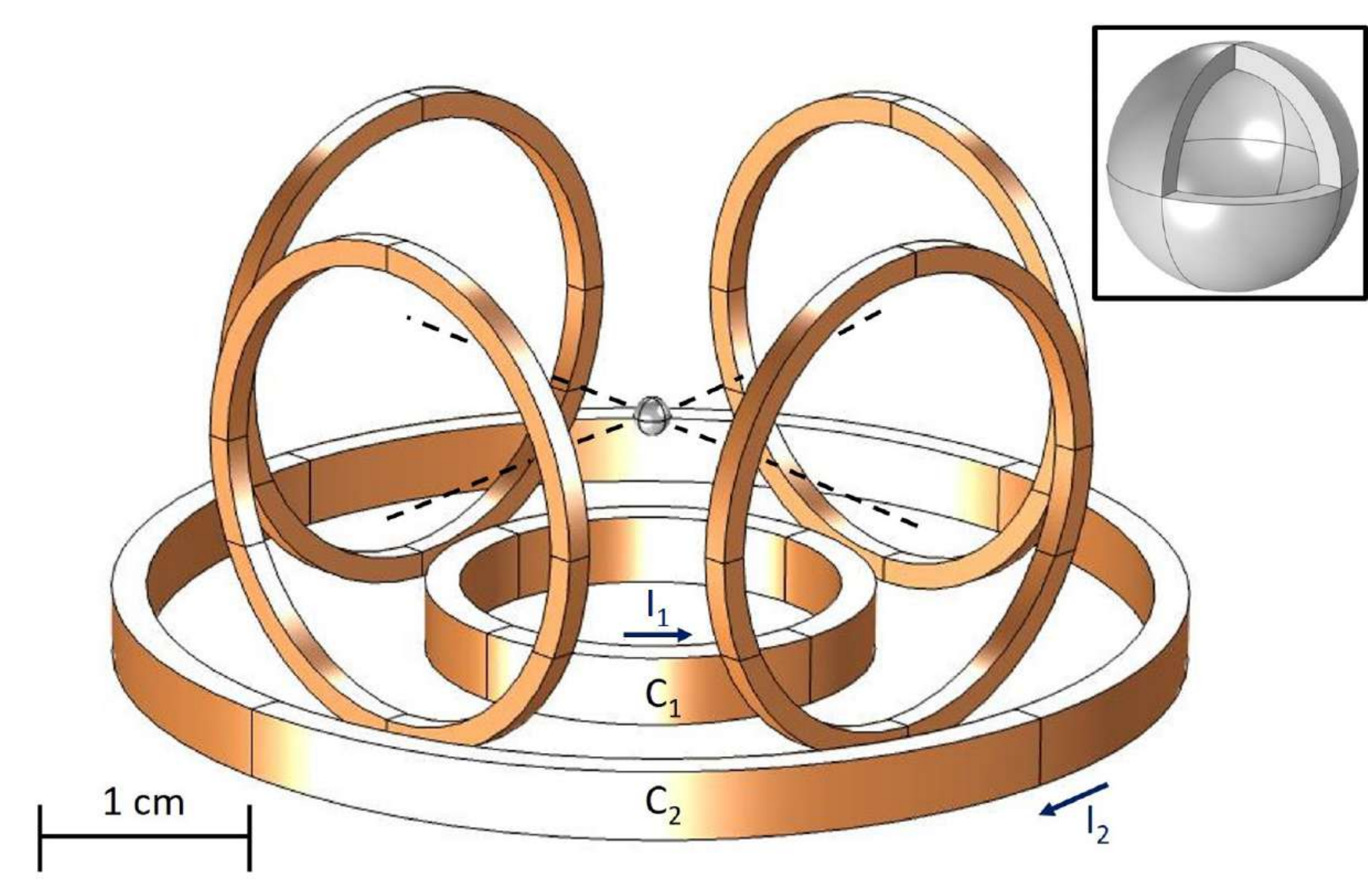}
  \caption{Experimental arrangement. The bottom coils ($C_1$, $C_2$) are fed with opposite currents. The lateral coils are placed in a symmetrical perpendicular configuration. The sphere is shown at the crossing of the axes of the lateral coils . The inset contains a detailed view of the hollow indium sphere.}\label{fig1}
  \end{center}
\end{figure}

Our experimental technique is based on the setup shown in Fig.~\ref{fig1}. Six superconducting coils are placed around the experimental cell, providing the levitation force and motion control of a superconducting sphere inside the cell. The cell, 12\,mm in diameter, is filled with superfluid $^3$He at temperatures below 200\,$\mu$K and contains vibrating wire bolometers for detecting energy losses resulting from the motion of the sphere in the superfluid. Two coils are placed at the bottom within a nested arrangement, with their symmetry axes on the vertical direction $z$, giving the main contribution to the levitating force. The four lateral coils placed on the sides have their axes pointing in the $x$ and $y$ directions, respectively. They have equal radii of 6\,mm and play the main role in the motion control of the sphere.   

 The inner bottom coil has a radius of $R_1=6$\,mm while the outer has the radius of $R_2=15$\,mm. We can create magnetic fields in opposite directions controlling the direction of the currents $I_1$ and $ I_2$ so we can obtain the right levitation conditions as we will discuss in the next section.

\subsection*{Stability}
A superconducting body exhibits diamagnetic behaviour. Diamagnetic objects are repelled by magnetic fields and can be levitated. However, in order to achieve stable levitation, the total energy must posses a minimum at the critical point, where gravitational and magnetic forces balance out. The stability conditions, as will be shown below, impose certain constraints on our design. On the other hand, we have limitations for the maximum currents in the coils, dictated by the critical current, and for the maximum field, dictated by the critical field of the superconducting material of the levitating object, which is 28\,mT in indium.

The differential amount of work done by the magnetic field on the body is $-d{\bf M}\cdot {\bf B}$,  while the induced magnetic moment reads \cite{Landau}
\begin{equation}
  \label{eq:moment}
  {\bf M}({\bf r})= \frac{\alpha\chi V{\bf B}({\bf r})}{\mu_0}.
\end{equation}
Note that the factor $\alpha$ appears as a geometric factor and $V$ is the volume of the body. In the case of a superconducting spherical body \cite{London} when the radius is bigger than the penetration length a good approximation is $\alpha=1/2$. Due to the perfect diamagnetic behaviour, the magnetic susceptibility is $\chi=-1$. Integrating the magnetic work as the field increases from zero to its final value, and taking into account the gravitational energy, we get the total energy 
\begin{equation}
  \label{eq:energy}
  U({\bf r}) = mgz + \frac{1}{4}\frac{VB^2({\bf r})}{\mu_0}.
\end{equation}
where $m$ is the mass of the body and $g$ is the gravitational acceleration. The points where levitation occurs are called critical points, as we have $\nabla U = 0$.
For the equilibrium to be stable, the energy must be at a minimum, i.e. the energy will increase in all the directions from the critical points. In mathematical form, the conditions are \cite{Berry97}
\begin{equation}
  \label{eq:mincond}
  \partial^2_xU>0, \;\;\; \partial^2_yU>0, \;\;\;\; \partial^2_zU>0.
\end{equation}

 We can perform an exact analysis of the stability. Up to small corrections, the field created by a coil can be approximated by the one created by a ring of radius $a$ with $n$ turns carrying a current $I$ as \cite{Landau}
\begin{equation}
       \label{eq:B_r}
     \begin{split}  
       B_r = \frac{\mu_0}{4\pi}\frac{2nIz}{r\sqrt{(a+r)^2+z^2}}\left(-K(k)+\frac{a^2+r^2+z^2}{(a-r)^2+z^2}E(k)\right)
     \end{split}
\end{equation}

\begin{equation}
      \label{eq:B_z}
       B_z=\frac{\mu_0}{4\pi}\frac{2nI}{\sqrt{(a+r)^2+z^2}}\left(K(k)+\frac{a^2-r^2-z^2}{(a-r)^2+z^2}E(k)\right)
\end{equation}
with $k=\sqrt{4ar/[(r+a)^2 + z^2]}$ as the argument for the elliptic integrals of the first and second type
\begin{equation*}
       K(k)=\int_0^{\frac{\pi}{2}} \frac{d\theta}{\sqrt{1-k^2\sin^2{\theta}}}, \;\;\; E(k)=\int_0^{\frac{\pi}{2}}\sqrt{1-k^2\sin^2{\theta}}\, d\theta,
\end{equation*}
and $B_r, B_z$ the radial and vertical components of the field.

Using this expression we can compute analytically the magnetic field created by the bottom coils.

The stability conditions (\ref{eq:mincond}) depend only on the magnetic field configuration as we can see from  (\ref{eq:energy}). Thus, in order to achieve a stable  levitation, we need to fulfill the condition $\nabla U =0$, in the region where
$\partial^2_x B^2>0, \;\;\; \partial^2_y B^2 >0, \;\;\;\; \partial^2_z B^2>0$,  here $B$ is the modulus of the magnetic field. An important consequence of the critical condition of the force being null is that it involves only the density $\rho$ of the levitated body as it reads
\begin{equation}
  \label{eq:critcond}
  B\partial_z B = -4\mu_o\rho g
\end{equation}
In order to control the density, the spherical body can be made by coating a plastic material with indium as represented in the inset of Fig.~\ref{fig1}. The coating thickness must be bigger than the penetration length which for indium is 60\,nm~\cite{Lock51}.  

\begin{figure}
  \begin{center}
  \includegraphics[width=0.65\textwidth]{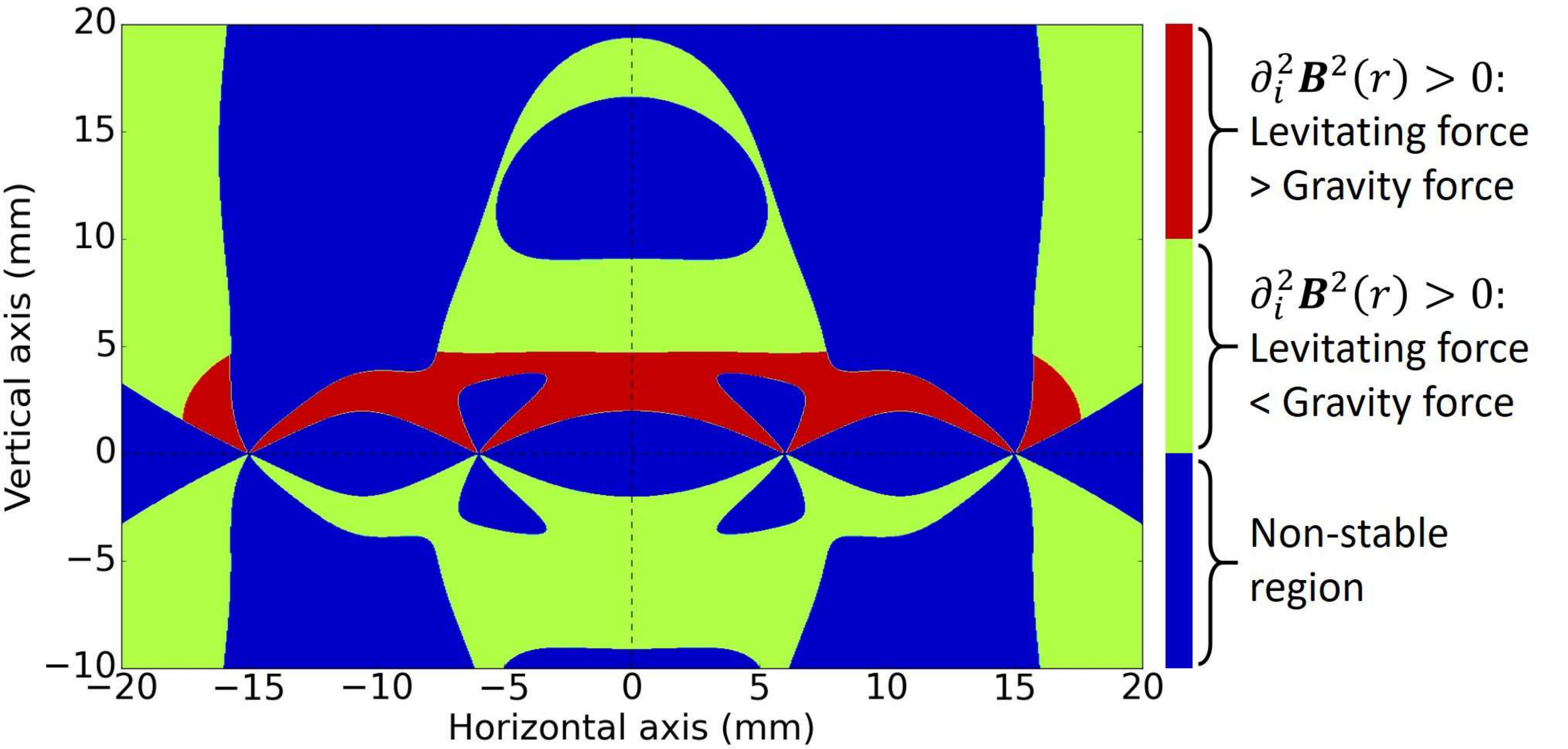}\\
  \caption{Stability/instability regions and critical line in a vertical cross section plane. Blue represents the unstable region, while red and green represent regions where the second derivative of the energy is greater than zero. The boundary between red and green regions is the zero force line, i.e the levitating position.\label{stability}}
  \end{center}
\end{figure}

In Fig.~\ref{stability} we have plotted again a vertical cross section showing the different stability regions when only the bottom coils are switched on. The parameters used are $n=250$ turns for both coils, $R_1= 6$ mm, $I_1= 2$ A for the inner coil, and $R_2= 1.5$ mm, $I_2= 1.7$ A (in the opposite direction) for the outer coil. The superconducting body is a hollow sphere made of indium, with internal radius $R_{in} = 500\, \mu\mbox{m}$, external radius $R_{out} = 550\, \mu\mbox{m}$ and a total mass of $1.26$ mg.  We show the unstable region in blue, while green and red represent the stable region, that is the region where $\partial^2_i B^2> 0$ . The red region is where the magnetic force in the $z$ direction is larger than the gravitational force, while in the green region the gravitational force is larger. Thus, the interface between the green and the red regions is the critical boundary, i.e. the zero force surface. In other words, the levitated body in equilibrium will be positioned at that interface. We observe that close to the centre there is an almost horizontal flat region of stable levitation of about 14 mm which is about 13 times the diameter of the sphere, providing the possibility to control the motion of the sphere in the horizontal direction with the lateral coils. 

\subsection*{The controlled motion of the sphere}
When the sphere is displaced out of the equilibrium position, it starts to oscillate. As the viscosity is very low, we need to suppress these oscillations by introducing a control system. The control system also will allow us to move the sphere with a constant velocity, which is crucial for the studies of surface excitations in $^3$He-B.

The motion around the equilibrium position in the horizontal direction are nonlinear harmonic oscillations whose frequency depends on the amplitude of the motion, but also on the intensity of the magnetic field produced by the bottom coils.  In order to prevent the oscillations and to move the sphere at a (quasi-)constant velocity, we can apply a pre-calculated force \cite{Zmeev14}, by running  a controlled current in the lateral coils.

As an example, in order to move the sphere from $x_1=0$\,mm to $x_2=+2$\,mm, with a uniform velocity of $v_0=0.02$ m/s we need to use a time-dependent current in the lateral coils of the form:
\begin{equation}
  \label{eq:control}
  I_L(t) = \left\{ \begin{array}{lcc}
  -389983t^4 +  7800t^3 - 15120t^2 + 151.2t & : & t \leq t_1 \\
  \\0.78t -0.004 & : & t_1 < t \leq t_2\\
  \\331485(t_2+t_1-t)^4 - 6630(t_2+t_1-t)^3 + 12852(t_2+t_1-t)^2 - 128.5(t_2+t_1-t) + 0.08 & : & t_2 < t \leq t_2 + t_1
  \end{array}
  \right.
\end{equation}

\begin{figure}
  \begin{center}
    \includegraphics[width=0.65\textwidth]{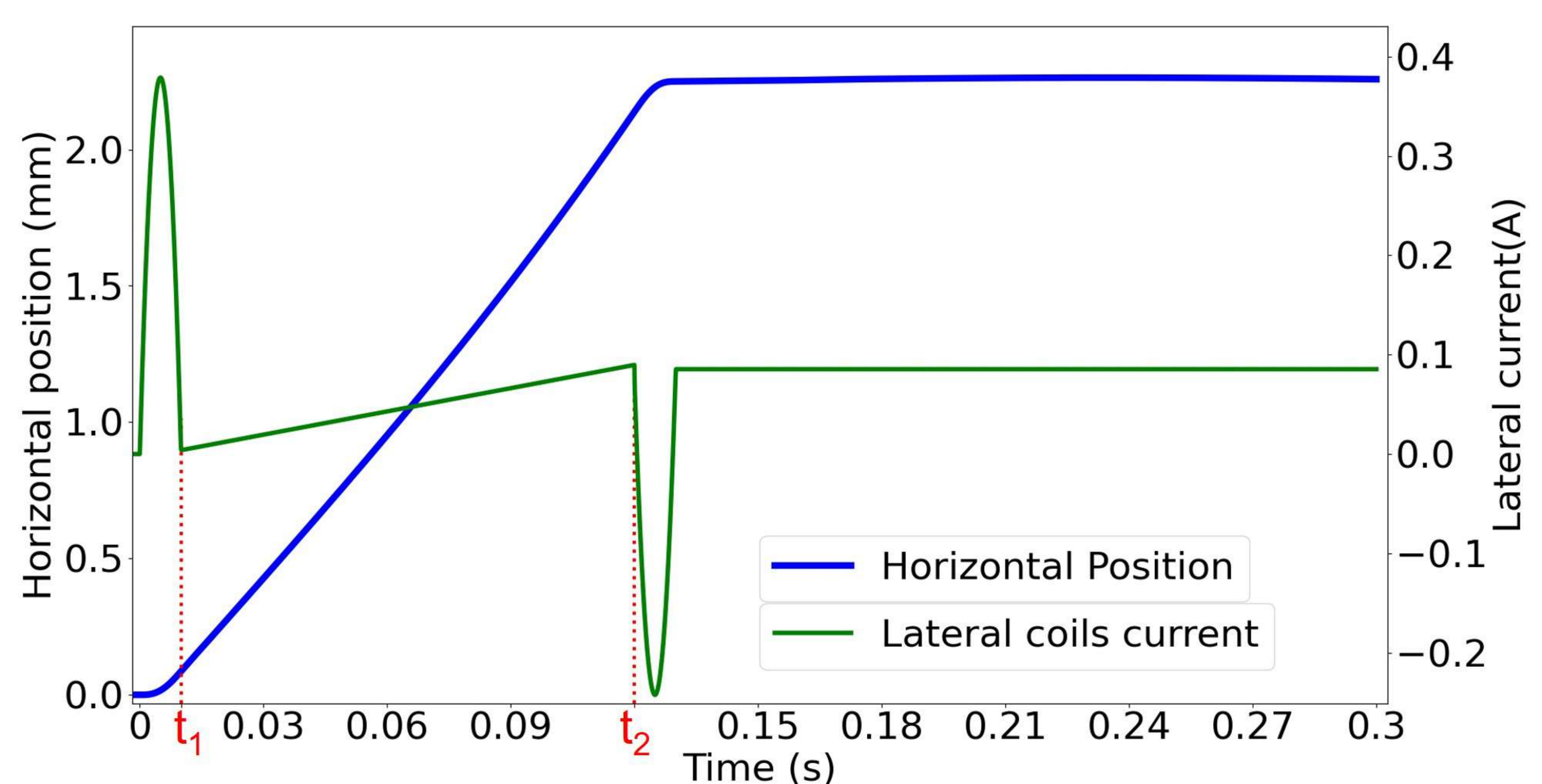}\\
  \caption{The controlled motion of the sphere in the horizontal plane. The motion of the sphere along the horizontal direction  is plotted in blue (thicker line),  the left axis being the displacement of the sphere. The green line represents the driving current applied to the lateral coils to induce the rectilinear movement, the magnitude of the current is indicated on the right axis.}\label{controlh}
  \end{center}
\end{figure}

Here the current $I_L(t)$ is in amps and $t$ is in seconds. The motion comprises three stages: acceleration, uniform motion and deceleration corresponding to the three time intervals in Eq.~\ref{eq:control}. The first stage, from $t=0$\,s to $t_1=0.01$\,s, represents the acceleration period of time for reducing the oscillations. The second stage, from $t_1$ to $t_2$, is the time interval in which uniform motion is induced, where $t_2=0.12$ s. The third stage, from $t=t_2$ to $t=0.13$ s, represents the time interval where the decelerating pulse is applied to stop the motion of the sphere. The lateral coils are fed with currents producing the field in the same direction, i.e. considering the coils in the y-axis, looking in the positive direction, both coils would carry clockwise currents to move the sphere in the positive $y$ direction. The result can be seen in Fig.~\ref{controlh}.

\begin{figure}
  \begin{center}
    \includegraphics[width=0.65\textwidth]{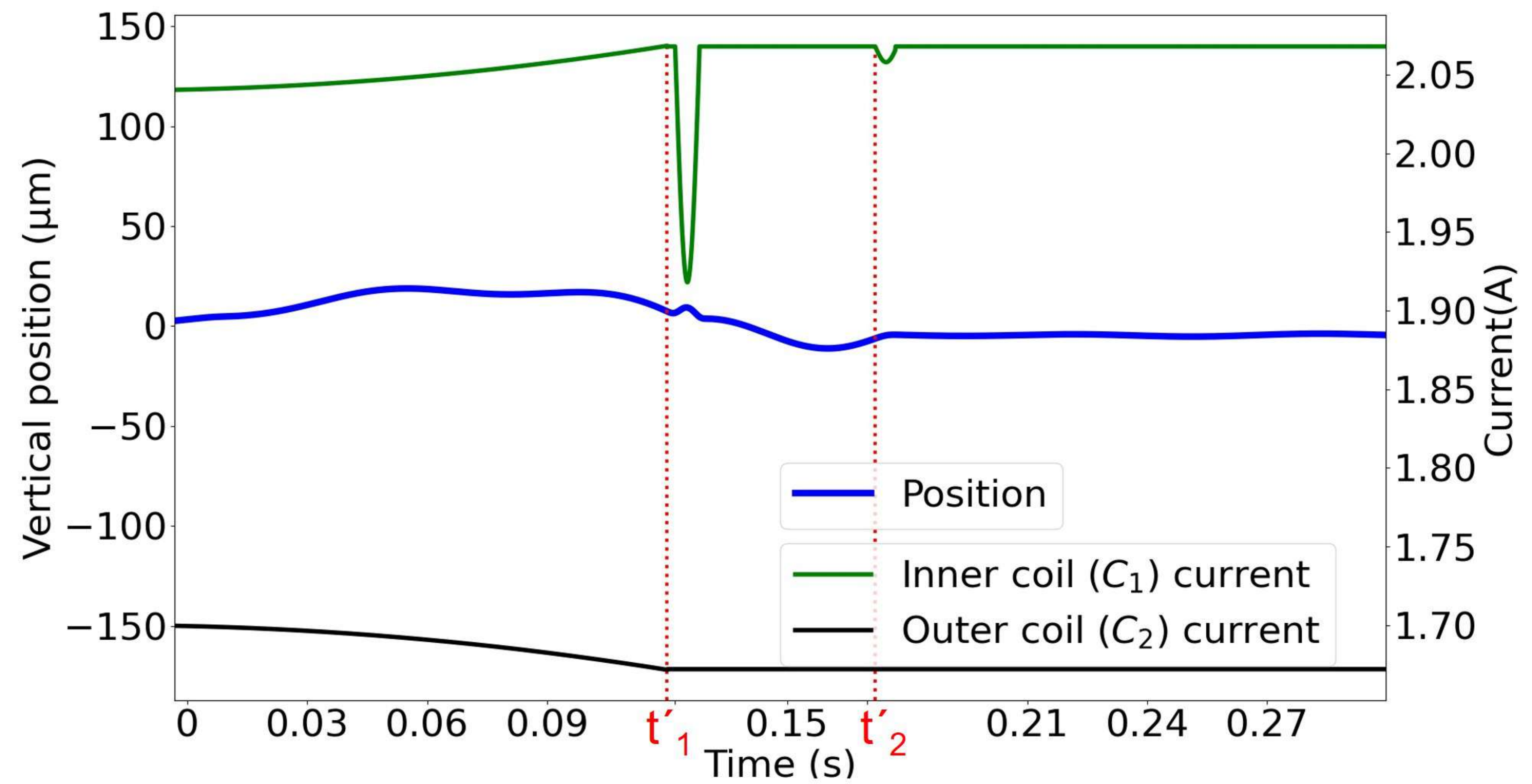}
    \caption{The controlled motion of the sphere in the vertical direction. The motion of the sphere along the vertical direction  is plotted in blue (thicker line). The green line (upper line) represents the current applied to the inner bottom coil ($C_1$), while the black line is the absolute value of the current applied to the outer bottom coil ($C_2$). Note that the currents in the coils have opposite directions. Counterphase pulses to decrease the magnitude of the oscillations are applied to the inner coil.\label{controlv}}
  \end{center}
\end{figure}

We can also use a similar control to suppress the vertical oscillations. These oscillations are produced due to the inhomogeneity of the magnetic field in addition to the distortion of the stability region produced by the side coils. The vertical control is implemented by adding a small synchronised time-dependent current to the bottom coils. The current in the inner coil is shown in Eq.~\ref{eq:controlv1} and the current applied to the outer coil is shown in Eq.~\ref{eq:controlv2}.  We can see in Fig.~\ref{controlv} that the oscillations are also removed in the vertical direction.

\begin{equation}
\label{eq:controlv1}
I_{C_{1}}(t) = \left\{ \begin{array}{lcc}
2.04 + 1.45t^2& : & t \leq t'_1 \\
\\2.068 - 0.14\sin(515(t-t'_1)) & : & t'_1 < t \leq t'_1 + 0.006\\
\\2.068 & : & t'_1 + 0.006< t \leq t'_2\\
\\2.068 - 0.01\sin(578(t-t'_2)) & : & t'_2 < t \leq t'_2 + 0.006\\
\\2.068 & : & t > t'_2 + 0.006
\end{array}
\right.
\end{equation}

\begin{equation}
\label{eq:controlv2}
I_{C_{2}}(t) = \left\{ \begin{array}{lcc}
1.7 - 1.45t^2& : & t \leq t'_1 \\
\\1.672 & : & t > t'_1
\end{array}
\right.
\end{equation}

The vertical control process consists of three different stages. In the first stage, from $t=0$ s to $t'_1=0.12$ s, the current is increased in the inner bottom coil ($C_1$) to compensate the lack of lift produced by the introduction of the horizontal control. Meanwhile, the magnitude of the current in the outer coil ($C_2$) is reduced to increase the upward magnetic field component. Without this stage, the vertical position would fall 200 $\mu$m, producing large oscillations in the process. In the second stage, at $t=t'_1$ a downward harmonic pulse (0.13 A in amplitude, 0.006 s long) is applied in the inner coil ($C_1$) to reduce the oscillations produced by the horizontal control at $t=t_2$.  These oscillations would produce increments in the horizontal velocity up to 2\,cm\,s$^{-1}$ without the control pulse. Finally,  at $t=t'_3=0.17$ s, another downward pulse (0.01\,A in amplitude, 0.006\,s long) is introduced in the inner coil ($C_1$) to completely remove the oscillations.

\subsection*{Detection of the sphere position}
We also need to implement the position detection system to monitor the motion of the sphere and to confirm the validity of our design. For this purpose we will utilise further two coils with axes along $x$ and $y$. These detection coils will be run at an AC current with a frequency by far exceeding all characteristic mechanical frequencies. The change in the position of the diamagnetic sphere will affect the inductance of the coils which will be registered as a change in their impedance.

\section*{Discussion}
As a proof of concept, we have developed a levitation and control system to explore properties of superfluid $^3$He at microkelvin temperatures. We will use the system as a mechanical probe. Our design allows us to move a body inside the superfluid in a wide range of motions, from oscillatory to uniform with the same setup. Next stage will imply the building of a working prototype.

The design consists of a pair of nested coils at the bottom of the cell, which provide the levitation force. The same coils play a role in the control of the sphere's motion. Two pairs of opposing lateral coils in a perpendicular arrangement complete the control system.  As shown in this work, the control system allows the uniform motion of the sphere in the horizontal plane.
The work presented includes a detailed study of the stability of the set-up and the motion of the sphere in the horizontal plane at uniform velocity. We include the time dependence of the current in the coils needed to control the spurious oscillations of the sphere. 

We believe that this levitronic device will open up a new trend in the low temperature research and the exploration of the mechanical properties of quantum fluids as well as topological defects within superfluids.   

\section*{Acknowledgements}
This work was funded by UK EPSRC (grant No. EP/P024203/1), the EU H2020 European Microkelvin Platform (Grant Agreement 824109) and by NATO Science for Peace and Security Programme (Secure Communication in the Quantum Era, G5448). 

\bibliography{bibliopaper}

\section*{Author contributions statement}
All the authors participated in the development of the design presented in the manuscript. The stability analysis was made by the URJC group. C.U. prepared the figures and ran the simulations. All the authors reviewed the manuscript. 

\section*{Additional information}
\textbf{Competing interests:} The authors declare no competing interests.

\end{document}